\documentclass[sigconf]{acmart}

\usepackage[english]{babel}
\usepackage[utf8]{inputenc}
\usepackage{blindtext}
\usepackage{booktabs} 
\usepackage{pdftexcmds}
\usepackage{graphicx}
\usepackage{subcaption}
\usepackage{rotating}
\usepackage{tikz}
\usetikzlibrary{external}
\usepackage{pdfpages}
\usepackage{pgfplots}
\usetikzlibrary{pgfplots.statistics}
\usepgfplotslibrary{groupplots}
\usepackage{wrapfig}
\usepackage{xcolor}
\usepackage{todonotes}

\pgfplotscreateplotcyclelist{my black white}{%
solid, every mark/.append style={solid, fill=gray}, mark=*\\%
dotted, every mark/.append style={solid, fill=gray}, mark=square*\\%
densely dotted, every mark/.append style={solid, fill=gray}, mark=otimes*\\%
loosely dotted, every mark/.append style={solid, fill=gray}, mark=triangle*\\%
dashed, every mark/.append style={solid, fill=gray},mark=diamond*\\%
loosely dashed, every mark/.append style={solid, fill=gray},mark=*\\%
densely dashed, every mark/.append style={solid, fill=gray},mark=square*\\%
dashdotted, every mark/.append style={solid, fill=gray},mark=otimes*\\%
dashdotdotted, every mark/.append style={solid},mark=star\\%
densely dashdotted,every mark/.append style={solid, fill=gray},mark=diamond*\\%
}

\makeatletter
\pgfplotsset{
    boxplot/hide outliers/.code={
        \def\pgfplotsplothandlerboxplot@outlier{}%
    }
}
\makeatother

\definecolor{lightgray}{rgb}{0.83, 0.83, 0.83}
\definecolor{lightslategray}{rgb}{0.47, 0.53, 0.6}
\definecolor{bole}{rgb}{0.47, 0.27, 0.23}
\definecolor{camel}{rgb}{0.76, 0.6, 0.42}

\renewcommand\footnotetextcopyrightpermission[1]{} 

\copyrightyear{2018}
\acmYear{2018}
\setcopyright{none}
\acmConference[CoNEXT '18]{The 14th International Conference on emerging Networking EXperiments and Technologies}{December 4--7, 2018}{Heraklion, Greece}
\acmBooktitle{The 14th International Conference on emerging Networking EXperiments and Technologies (CoNEXT '18), December 4--7, 2018, Heraklion, Greece}
\acmPrice{15.00}
\acmDOI{10.1145/3281411.3281426}
\acmISBN{978-1-4503-6080-7/18/12}

\begin{document}
\title[eBPF for IPv6 Segment Routing]{Leveraging eBPF for programmable network functions with IPv6 Segment Routing}

\author{Mathieu Xhonneux, Fabien Duchene, Olivier Bonaventure}
\affiliation{%
  \institution{UCLouvain, Louvain-la-Neuve, Belgium}
}
\email{first.last@uclouvain.be}


\begin{abstract}
With the advent of Software Defined Networks (SDN), Network
Function Virtualisation (NFV) or Service Function Chaining (SFC), operators
expect networks to support flexible services beyond the mere forwarding of
packets. The network programmability framework which is being developed within
the IETF by leveraging IPv6 Segment Routing enables the realisation of in-network functions.

In this paper, we demonstrate that this vision of in-network programmability can be
realised. By leveraging the eBPF support in the Linux kernel, we implement a flexible
framework that allows network operators to encode their own network functions as eBPF
code that is automatically executed while processing specific packets. Our lab measurements
indicate that the overhead of calling such eBPF functions remains acceptable. Thanks to eBPF,
operators can implement a variety of network functions. We describe the architecture of our
implementation in the Linux kernel. This extension has been released with Linux 4.18.
We illustrate the flexibility of our approach with three
different use cases: delay measurements, hybrid networks and network discovery. Our lab measurements
also indicate that the performance penalty of running eBPF network functions on Linux routers does not incur a significant overhead.
\end{abstract}

\maketitle

\section{Introduction}

In the late 1990s, various researchers proposed active network
architectures where each packet could carry code that would be
executed by intermediate routers while packets are forwarded to their
final destination \cite{tennenhouse1996towards}. Several of these
architectures were prototyped
\cite{tennenhouse1997survey,smith1999activating} but none was
deployed \cite{calvert2006reflections}. Still, these efforts were
precursors for middleboxes \cite{feldmeier1998protocol} and Software
Defined Networks (SDN) or Network Function Virtualisation
(NFV). SDN enables network operators to better control the flow of
packets in their networks 
\cite{mckeown2008openflow,kreutz2015software}. It
has been mainly targeted at enterprise networks.
NFV aims
at enabling network operators to deploy specialised network functions which can process
all the packets for specific flows. NFV is interesting for both enterprise and ISP
networks. A frequently cited use case for NFV are the future 5G networks that will need 
to combine a variety of functions.

Segment Routing  \cite{filsfils2015segment}, initially proposed as a simplification of
MultiProtocol Label Switching (MPLS) for ISP networks has gradually evolved into a much
more generic solution. Segment Routing is a modern version of source routing. It
enables routers to forward packets along a succession of shortest paths, each of them
being identified by a segment. Besides the MPLS variant, the IETF is currently developing an
IPv6 variant of Segment Routing (SRv6) that is gaining a lot of interest \cite{sr6demo,NANOG71}. 
Coupled with the fast deployment of IPv6 during the last years, this opens new opportunities
for both ISP and enterprise networks.

SRv6 leverages the flexibility of the IPv6 packet format and the large IPv6 addressing space. 
With SRv6, each segment is encoded as an IPv6 address that is advertised through the
intradomain routing protocol. NFV can be supported by assigning a specific address to each
network function and using segments to forward specific flows towards the appropriate
functions. 

This paper focuses on the NFV use case with SRv6. We extend the SRv6 implementation
in the Linux kernel \cite{LB17} to support the ability to run specific network functions on 
a per-packet basis. To allow more flexibility compared to other solutions like 
P4~\cite{p4}, our implementation leverages the eBPF support of the Linux kernel 
to
enable network operators to write their own intra-domain network functions that are dynamically linked to
the Linux kernel. Our evaluation shows that this enables the network functions
to run efficiently inside the kernel. We then demonstrate three use cases showing 
very different network functions as an illustration of the flexibility of our approach.

\section{Background}\label{sec:background}

Segment Routing \cite{filsfils2015segment} started as a modern variant of the source routing paradigm \cite{RFC8402} using the MPLS dataplane. 
This architecture has now evolved to also encompass the IPv6 dataplane \cite{I-D.ietf-6man-segment-routing-header}. Segment Routing in the IPv6 data plane (SRv6) is implemented by adding
an IPv6 extension header called the Segment Routing Header (SRH)
\cite{I-D.ietf-6man-segment-routing-header}. This SRH contains
one or more 128-bits IPv6 addresses that encode the \textit{segments}, the nodes that must be visited on the path
between the source and the destination. In the first versions
of the IPv6 variant of Segment Routing, these addresses were used
to identify routers and outgoing links. Several implementations of SRv6 have been announced, on commercial routers \cite{sr6demo} and on Linux \cite{LB17,abdelsalamsera}.

In 2017, the idea of SRv6 network programming emerged \cite{srv6-network-programming}. It generalises the 
notion of segments. Each path specified by a source 
is decomposed into an ordered list of instructions,
called \textit{segments}. Each segment, or endpoint, represents a \textbf{function} to be
executed at a specific location in the network. 
These functions may range from simple
topological instructions (e.g.\ forwarding a packet on a specific link) to more complex
user-defined behaviours. The SRH carries the ordered list of segments in each packet and
optionally Type-Length-Value (TLV) fields. 
The TLVs are 3-tuples that can be used to store additional data in
the SRH, e.g.\ for OAM purposes. 

The basic processing of packets with a SRH is the endpoint \texttt{End} function,
which advances the SRH to the next segment and forwards the packet to the destination corresponding to the segment
\cite{srv6-network-programming}. Several other functions extend this processing, such as \texttt{End.X},
which after advancing to the next segment, forwards it to a specific IPv6 nexthop, \texttt{End.T},
which performs the lookup for the next segment in a IPv6 routing table bound 
to the segment,
\texttt{End.B6}, which inserts a new SRH on top of the existing one, etc.  SRv6 
actions can also be applied on packets
without a SRH, e.g.\ inserting an SRH in an IPv6 packet, and encapsulating an 
outer IPv6 header with a SRH.
These actions are called transit behaviours. Between segments, packets are forwarded along the shortest path. 

The SRv6 implementation in the Linux kernel \cite{LB17,abdelsalamsera} supports the
basic features of SRv6 on hosts (sending and receiving IPv6 packets with an SRH) and on routers (forwarding SRv6 packets) but not all the
recent SRv6 network programming features. The current Linux SRv6 implementation uses two lightweight tunnels, \texttt{seg6} and \texttt{seg6local} to support the basic SRv6 mechanisms which can be plugged in the IPv6 layer.
\texttt{seg6} allows to implement the two transit behaviours mentioned above, i.e.\ inserting or
encapsulating SRHs in traffic matching a given destination, whereas \texttt{seg6local} allows
an operator to install SRv6 segments mapped to specific SRv6 functions, along with the required parameters. 
The set of actions provided by \texttt{seg6local} is bounded to a few simple
functions statically implemented in the kernel. They do not enable extensive
network programming capabilities, as required by SDN \cite{lebrun2018software},
SFC \cite{abdelsalam2017implementation,srv6pipes} and other
\cite{desmouceaux2017srlb,sr-frr,I-D.spring-srv6-oam}
applications in the SRv6 data plane.

\subsection{eBPF, an in-kernel virtual machine}

eBPF (for \textit{extended Berkeley Packet Filter}), is a general-purpose virtual machine that is included in the Linux kernel since the 3.15 release.
This virtual machine supports a 64 bits RISC-like CPU \cite{bpf-doc-kernel} 
which is an extension of the BPF virtual machine \cite{mccanne1993bsd}. 
It provides a programmable interface to adapt 
kernel components at run-time to user-specific behaviours. 
While solutions such as~\cite{pisces,silkroad} use P4~\cite{p4} to achieve data 
plane programmability, they are limited by the fact that P4 relies on specific 
hardware (and/or compiler), while eBPF targets a general purpose CPU and can be 
used on devices like Customer-Premises Equipment (CPE).
The LLVM project \cite{llvm} includes a BPF backend, capable of compiling C programs to BPF bytecode.
eBPF bytecode is either executed in the kernel by an interpreter or translated to native machine code using a Just-in-Time (JIT) compiler.
Since the eBPF architecture is very close to the modern 64-bit ISAs, the JIT compilers usually produce efficient native code \cite{bpf-lwn}.

eBPF programs can be attached to predetermined hooks in the kernel. Several hooks are available
in different components of the network stack, such as the traffic classifier (\texttt{tc})
\cite{borkmann2016getting}, or the eXpress Data Path (XDP) \cite{bpf-xdp}, a low-level hook
executed before the network layer, used e.g.\ for DDoS mitigation.
When loading an eBPF program into the kernel, 
a verifier first ensures that it cannot threaten the stability and security of 
the kernel (no invalid memory accesses, possible infinite loops, \ldots). The eBPF program is then executed
for each packet going through the datapath associated to its hook. The program can read and, for some hooks, modify the packet.

eBPF programs can call \textit{helper functions} \cite{bpf-helpers}, which are 
functions implemented in the kernel.
They act as proxies between the kernel and the eBPF program. Using such helpers, eBPF programs can retrieve and push data from or to the kernel,
and rely on mechanisms implemented in the kernel. A given hook is usually associated with a set of helpers. 

There are two practical issues when developing eBPF programs. The first is how to store persistent state and the second is how it can communicate with user space applications. State can be kept persistent between multiple eBPF program invocations and shared with user space applications using \textit{maps}.
Maps are data structures implemented in the kernel as key / value stores \cite{bpf-xdp}. Helpers are provided to allow eBPF programs to retrieve and store data into maps.
Several structures are provided, such as arrays, hashmaps, longest prefix match tries, \ldots When processing packets, if information needs to be pushed asynchronously
to user space, \textit{perf events} can be used. Perf events originate from Linux's performance profiler \texttt{perf}.
In a networking context, they can be used to pass custom structures from the eBPF program to the perf event ring buffer along with the packet being processed \cite{bpf-perf-events}. 
The events collected in the ring buffer can then be retrieved in user space. These mechanisms allow stateful processing and a user-space communication that would be difficult to achieve with P4.

Finally, a lightweight tunnel infrastructure named \textit{BPF LWT}, provides generic hooks in several network layers, including IPv6 \cite{bpf-lwt}. This LWT enables the execution of eBPF programs
at the ingress and the egress of the routing process of network layers, but is unable to leverage the specificities of SRv6.

\section{The SRv6 eBPF interface}\label{sec:interface}

Many of the emerging concepts of network functions leveraging SRv6
\cite{srv6-network-programming} cannot be deployed using the static 
actions that are supported by the existing \texttt{seg6local} infrastructure
on Linux \cite{LB17}. Adding explicit support for each SRv6 network
function in the Linux kernel would be difficult since the set of
functions continues to evolve. A better approach would be to include
in the Linux kernel a set of generic functions that allow network
operators to implement their own SRv6 functions. For this, we propose
a new eBPF interface to efficiently implement a broad
range of SRv6 actions.

Our hook is a new action \texttt{End.BPF} in \texttt{seg6local}. 
Each instance of this action is bound to an eBPF program.
It behaves as an endpoint, i.e.\ it only accepts SRv6 packets with a current segment
corresponding to a local eBPF action, advances the SRH to the next
segment, and subsequently executes the associated eBPF code.
We have designed this action with two key principles in mind: $(i)$ 
eBPF code cannot compromise the stability of the kernel and
$(ii)$ eBPF code should be able to leverage all the functionalities of
the SRv6 data plane.

To guarantee the stability, we need to ensure that \texttt{End.BPF}
can only allow write access to fields of the packet which can be modified by SRv6 endpoints.
The \texttt{seg6local} actions are executed in the IPv6 layer,
and further processing of the packet after \texttt{End.BPF} 
requires the packet to be valid.
Instead of providing direct-write access to the packet to the eBPF
code, we provide a specific helper function that restrains 
the fields which can be modified. This function also checks whether
any modification to these fields could jeopardise the integrity of the SRH.

\subsection{SRv6 API for eBPF programs}

All SRv6 eBPF programs are called with the packet as argument. They
have full read access to its payload, starting from the 
outermost IPv6 header. We designed three SRv6 specific helpers 
to extend the functionalities of our interface:

\begin{itemize}
    \item \texttt{bpf\_lwt\_seg6\_store\_bytes}: provides indirect write access
        to the editable fields of the SRH (i.e.\ the flags, the tag, and 
        the TLVs).
    \item \texttt{bpf\_lwt\_seg6\_adjust\_srh}: allows growing or shrinking the space reserved to TLVs.
    \item \texttt{bpf\_lwt\_seg6\_action}: executes a basic SRv6 function.
        It provides access to \texttt{End.X}, \texttt{End.T}, \texttt{End.B6}, \\
        \texttt{End.B6.Encaps} and \texttt{End.DT6}.
\end{itemize}

All SRv6 eBPF programs return an integer. This return value decides
the subsequent processing of the packet. Three values can be returned:
\begin{itemize}
    \item \texttt{BPF\_OK}: a regular FIB lookup must be performed on the next segment,
        the packet must be forwarded on the egress interface returned by the lookup.
    \item \texttt{BPF\_DROP}: the packet must be dropped.
    \item \texttt{BPF\_REDIRECT}: the default endpoint lookup must not be performed, and the packet must be forwarded to the destination already set in the packet metadata. 
\end{itemize}

The structure containing the packet that is passed to an eBPF program
contains both the payload and metadata such as its destination. 
When \texttt{bpf\_lwt\_seg6\_action} is called with
an action requiring a FIB lookup (e.g. \texttt{End.X}), 
the helper performs the requested FIB lookup and stores the result
in the metadata. When the execution of the program finishes, 
it is important that \texttt{End.BPF} does not execute the default 
lookup to the next segment afterwards, otherwise the destination 
previously set would be overwritten, hence the need for \texttt{BPF\_REDIRECT}.
If the SRH has been altered by the BPF program, a quick verification is
performed to ensure that it is still valid (e.g. if the SRH has grown, 
ensure that the allocated space has been filled with valid TLVs), otherwise it is dropped.
Finally, the packet is yielded back to the IPv6 layer, which takes care of
the forwarding to the destination set in the metadata of the packet.

A fourth helper function, \texttt{bpf\_lwt\_push\_encap}, has been
implemented in the \texttt{BPF LWT} hook. It allows to insert an SRH or encapsulate 
an outer IPv6 header with an SRH in pure IPv6 traffic. 

Both \texttt{End.BPF} and the helpers have officially been included into the Linux kernel,
and effectively released since Linux 4.18.



%


\subsection{Performance evaluation}
\label{subsec:bpf-perfs}

\begin{figure}[ht!]
\centering
\includegraphics[width=\linewidth]{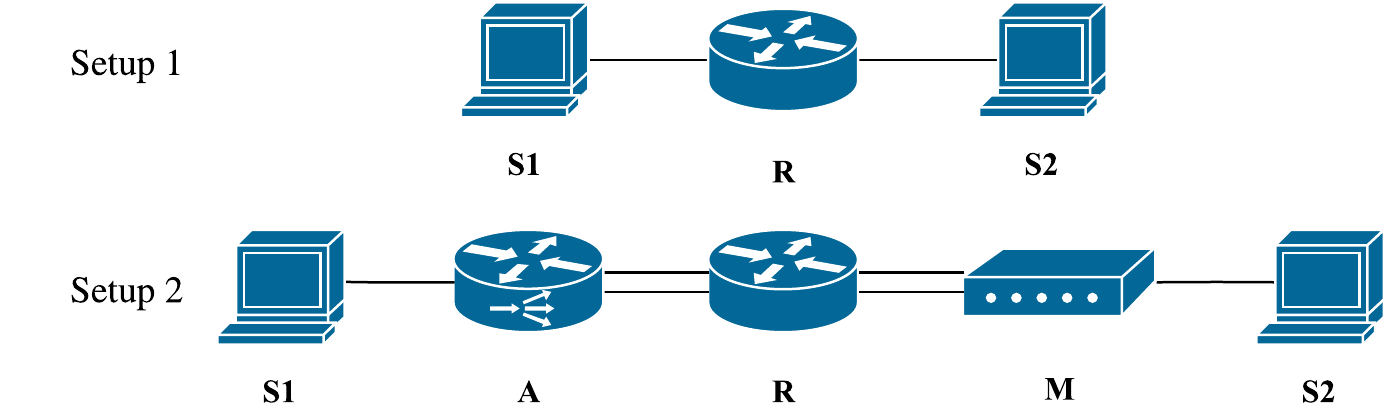}
\caption{Lab setups used in our experiments.}
\label{fig:setups}
\end{figure}

To measure the performance of our implementation of the \texttt{End.BPF} function,
we ran several measurement campaigns in a small lab.
Our lab is composed of 3 servers with Intel Xeon X3440 processors,
16GB of RAM and 10 Gbps NICs (setup 1 in Figure \ref{fig:setups}). 
Although these servers have multiple cores, we configured the
interrupts of their NICs to direct all received packets to the same
CPU core. 
We use \texttt{trafgen} to generate UDP packets on S1. Each UDP packet has a
payload of 64 bytes and an SRH with two segments, one bound to a function on R,
and the address of S2. R executes the endpoint functions while S2 acts as a
sink. 

For all experiments, unless stated otherwise, we enabled the JIT
compiler when running our eBPF code. We first measured the raw IPv6
packet forwarding performance with those UDP packets. When the source
sent 3 million packets per second, the rightmost server only received
them at a rate of 610 kpps. We use this number as the reference to
evaluate the impact of executing eBPF code while forwarding
each packet.


\begin{figure}[ht!]
	\centering
	\begin{tikzpicture}
	\begin{axis}[
	x tick label style={
		/pgf/number format/1000 sep=},
	ylabel={Packets forwarded per second, normalized},
	ymajorgrids,
	ymin=0,
	ymax=100,
	ybar,
	width=1\linewidth,
    height=6cm,
	enlarge x limits=0.05,
	bar width=15pt,
	xticklabel style={align=center},
	ytick={0,20,40,60,80,100},
	yticklabels={0\%,20\%,40\%,60\%,80\%,100\%},
	xtick={1,2,3,4,5,6,7},
	xticklabels from table={labels-simple-end.dat}{Label},
	]
	\addplot [lightgray!10!black,fill=bole]
	coordinates {(1,83.02288551) (2,80.46670325) (3,89.85687837) (5,77.77242772) (4,84.14655668) (6,75.15737807) (7,42.66801296) };
	\end{axis}
	\end{tikzpicture}
	\caption{Simple endpoint functions are efficiently supported.}
	\label{fig:kernels-perfs-simple}
\end{figure}
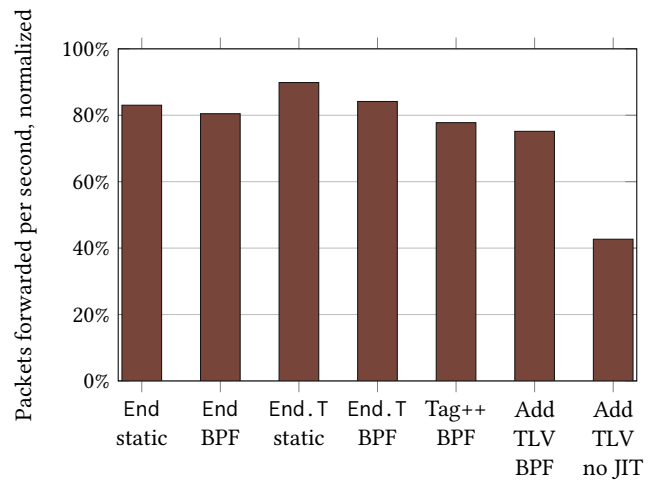

We evaluate four simple \texttt{End.BPF} functions. 
Our first eBPF code is \texttt{End}, i.e.\ a BPF function that does
nothing (1 source line of code (SLOC) in its body).
This function serves as a baseline to evaluate the cost
of calling an eBPF function when forwarding each packet.
Our second function is a BPF counterpart for \texttt{End.T},
calling \texttt{bpf\_lwt\_seg6\_action} (4 SLOC).
Our third eBPF program, \textit{Tag++}, fetches the tag of the SRH and
increments it by doing an indirect write using \texttt{bpf\_lwt\_seg6\_store\_bytes} (50 SLOC). 
The last one, \textit{Add TLV}, adds an 8-byte TLV. This requires a
call to \texttt{bpf\_lwt\_seg6\_adjust\_srh}, followed by a call to \\
\texttt{bpf\_lwt\_seg6\_store\_bytes} to fill the newly allocated space
with the content of the TLV (60 SLOC).


Figure \ref{fig:kernels-perfs-simple} shows the relative forwarding
performance of these functions compared to raw IPv6 forwarding. Each
point is the average over 20 measurements. The normalised standard
deviation was below 1\%.
Compared to the static \texttt{seg6local} implementation directly written in
the kernel, the eBPF equivalent of \texttt{End} has a reduced throughput of
only 3\%. In addition, \textit{Tag++} decreases this throughput by 3\%, 
by fetching and incrementing the tag through one helper call.
The equivalent of \texttt{End.T} is capable of forwarding a 5\% smaller
throughput than its static counterpart.
Finally, \textit{Add TLV} forwards a 5\% lower throughput than
\texttt{End} written in BPF. \textit{Add TLV} and \textit{Tag++} do not have
static counterparts in the \texttt{seg6local} infrastructure.
In all four cases, the performance overhead is deemed acceptable.

Moreover, we use \textit{Add TLV} to evaluate the benefits of
using the eBPF JIT to support SRv6 network programming. When disabling the JIT
compiler, the throughput going through \textit{Add TLV} is divided by a factor
of 1.8. Similar factors have been observed when evaluating the impact of the JIT
compiler on other programs with similar complexities.
This factor is expected to increase when the number of instructions per BPF program also increases. 

\section{Use-cases}

By leveraging \texttt{End.BPF}, network operators can implement their 
own in-network functions that are applied on a per-packet
basis. As \texttt{End.BPF} is implemented on Linux, it can be used in 
a variety of environments. One important use case are the low-end
access routers that are placed in many homes. We illustrate the
flexibility of \texttt{End.BPF} by demonstrating three very
different use cases in this section. 

\subsection{Passive monitoring of network delays}

Delay is one of the most important performance metrics in a
network. Network operators use a variety of techniques,
ranging from simple \texttt{ping} to more precise measurements
\cite{rfc7679,ali-spring-srv6-pm}. While solutions such 
as~\cite{pingmesh,telemetry} have been proposed to measure the latency in 
datacenter networks, we propose a solution that can be deployed in the edge, up 
to the Customer-Premises Equipment (CPE). 
To illustrate \texttt{End.BPF},
we implement the recently proposed one-way delay (OWD) delay
measurement for SRv6 \cite{ali-spring-srv6-pm} through our eBPF interface.

Our solution uses two eBPF programs installed at both tips of the path being monitored.
On the router at the beginning of the path, a \texttt{BPF LWT} program is executed
for each packet towards the given destination. This program
encapsulates, using the \texttt{bpf\_lwt\_push\_encap} helper, 
a defined percentage (or \textit{probing ratio}) of the incoming regular IPv6 packets with an SRH.
This SRH contains a \textit{Delay Measurement} (DM) TLV, with a 64-bit 
timestamp inserted by the router that processed the packet,
and a second TLV containing the IPv6 address and UDP destination port 
of the controller that collects the delay measurements. 
The segment list enforces the path on which the delay is monitored.
The last segment corresponds to the router at the end of the monitored path, with the \texttt{End.DM} instruction.
The SRH is built by the BPF program, the transmission timestamp is
retrieved using a generic helper that we added to the Linux kernel.
This function is written in 130 SLOC.

\texttt{End.DM} is an SRv6 network function implemented using \texttt{End.BPF}.
At the beginning of its execution, it fetches the RX software timestamp, i.e.\ the time the packet left the
NIC driver and entered the kernel. It subsequently inspects the SRH to retrieve the TX timestamp inside the DM TLV,
as well as the TLV containing the address of the controller. Both timestamps and the information regarding the controller
are sent to a user space daemon using a \texttt{perf} event since an 
eBPF program is not capable of sending out-of-band replies.
Finally, it decapsulates the outer IPv6 header using \texttt{bpf\_lwt\_seg6\_action} with the \texttt{End.DT6} action,
and indicates that the inner IPv6 packet should be forwarded
normally.

Our user space daemon is implemented in Python,
it continuously listens for perf events. When an event is received, it
creates a new thread to send both timestamps to the indicated
controller in a single UDP datagram.
The implementation uses the \texttt{bcc} framework \cite{bcc}, a BPF front-end in Python
giving straightforward access to perf events, and is written in 100 SLOC.

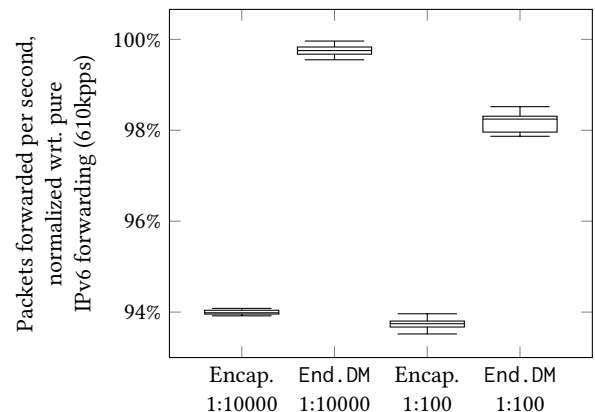
\begin{figure}[ht!]
\begin{tikzpicture}
    \begin{axis}[boxplot/draw direction=y, boxplot/hide outliers,
    compat=newest,
    width=0.85\linewidth,
    xticklabel style={align=center},
    xtick={1,2,3,4},
	xticklabels from table={labels-powd.dat}{Label},
    ymin=93,
    ytick={94,96,98,100},
    yticklabels={94\%,96\%,98\%,100\%},
    xlabel style={align=center, text width=8cm},
    ylabel style={align=center},
    ylabel={Packets forwarded per second,\\ normalized wrt. pure \\IPv6
            forwarding (610kpps)}]
            \addplot [boxplot, mark=*] table [col sep=semicolon, y=t10000] {4-otp-encap.csv};
            \addplot [boxplot, mark=*] table [col sep=semicolon, y=t10000] {4-otp-owd.csv};
            \addplot [boxplot, mark=*] table [col sep=semicolon, y=t100] {4-otp-encap.csv};
            \addplot [boxplot, mark=*] table [col sep=semicolon, y=t100] {4-otp-owd.csv};
    \end{axis}
\end{tikzpicture}
\caption{Impact of both BPF programs on the forwarding performances,
    for two probing ratios.}
\label{plot:otp-decap}
\end{figure}

We evaluated the performance impact of both BPF programs using the setup described in \ref{subsec:bpf-perfs}.
R executes the \texttt{End.DM} and transit behaviour eBPF programs. 
S1 uses \texttt{pktgen} to generate IPv6 packets
without SRH, and \texttt{trafgen} for packets with a DM TLV. 
The results are presented in Figure \ref{plot:otp-decap}.

These results show that our passive delay monitoring is executed with almost no impact on the forwarding performances,
even with a probing ratio of 1:100. 
We note that the transit behaviour forwards only 5\%
fewer packets than the native IPv6 datapath.
\texttt{End.DM} has virtually no impact on performances for a 1:10000 probing 
ratio,
even considering that all packets with a DM TLV are decapsulated.

\subsection{Hybrid Access Networks}


In order to offer higher bandwidths to their clients, ISPs have started to
deploy hybrid access networks \cite{tr348}, i.e.\ networks that combine
different access links such a xDSL and LTE. In one deployment, described in
\cite{RFC8157}, a hybrid CPE router with xDSL and LTE is connected to
an aggregation box with GRE Tunnels. The tunnels ensure that the
packets sent by the hybrid CPE are routed to the aggregation box that 
reorders them. Since SRv6 inherently allows controlling over which path 
each packet is forwarded, we wondered whether it is possible to
leverage SRv6 BPF to design and implement a link aggregation
solution that achieves good performance on home routers (CPE).

As \cite{tr348}, we use an aggregation box deployed in the ISP
backbone. For each IPv6 route towards a client, a \textit{LWT eBPF} program is
installed on the aggregation box. It encapsulates the
packets towards the client into an IPv6 header with an SRH. They are then
decapsulated upon reception by the CPE and forwarded to the destination in the
client's LAN. The SRv6 decapsulation is natively performed by the kernel.
The CPE uses the same eBPF program to
encapsulate its packets with an SRH towards the aggregation box.

The eBPF program leverages the \textit{LWT eBPF} hook and \\
\texttt{bpf\_lwt\_push\_encap} to encapsulate an SRH.
Our implementation, written in 120 SLOC, performs a per-packet Weighted
Round-Robin (WRR) scheduling to aggregate the bandwidths of two links.
The weights of the WRR match the uplink links capacities, as seen by the CPE
or the aggregation box. We use maps to store the scheduler state,
i.e.\ the weights and the last chosen path.

To experiment with such hybrid access networks, we first 
configured our network as the setup 2 in Figure\ref{fig:setups}.
S1, S2, A and R are the same servers as used in
\ref{subsec:bpf-perfs}. S1 and S2 acts as end hosts. 
Node A acts as the aggregation box. Our CPE, M,
is a Turris Omnia router
with a 1.6 GHz dual-core ARMv7 processor
and 1~Gbps NICs. It is recent and runs OpenWRT (hence easily modifiable),
making it an ideal subject for our use-case.
R uses \texttt{tc netem} to insert latency on the 
links and to limit their bandwidth.
We configure one link with a bandwidth of 50~Mbps, 
an average RTT of 30~ms and a standard deviation of 5~ms.
The other has a bandwidth of 30~Mbps, an average RTT of 5~ms 
with a standard deviation of 2~ms. These values mimic current
metrics of average broadband access networks.

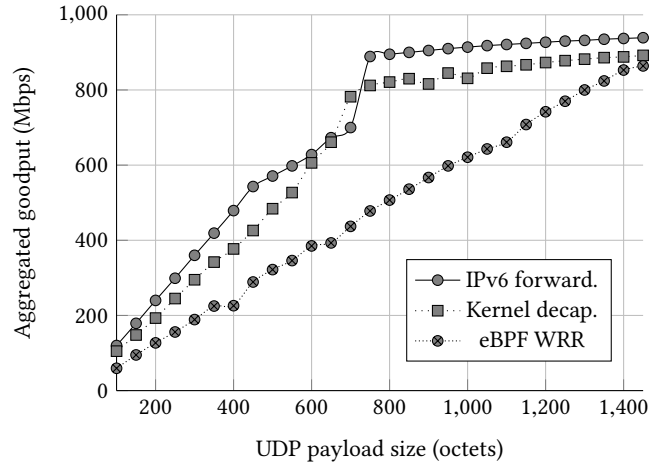
\begin{figure}[ht!]
\centering
\begin{tikzpicture}
\begin{axis}[
   width=7cm,
   height=5cm,
   scale only axis,
   xmin=100, xmax=1450,
   xmajorgrids,
   ymin=0, ymax=1000,
   xlabel={UDP payload size (octets)},
   ylabel={Aggregated goodput (Mbps)},
   ymajorgrids,
   axis lines*=left,
   legend style ={ at={(0.55,0.35)},
        anchor=north west, draw=black,
        fill=white,align=left},
    cycle list name=my black white,
    smooth
]
        \addplot table [
            col sep=comma,
            x=payload,
            y=baseline,
        ] {5-bw-pktsize.csv};
        \addlegendentry{IPv6 forward.};

        \addplot table [
            col sep=comma,
            x=payload,
            y=decap,
        ] {5-bw-pktsize.csv};
        \addlegendentry{Kernel decap.};

        \addplot table [
            col sep=comma,
            x=payload,
            y=encap,
        ] {5-bw-pktsize.csv};
        \addlegendentry{eBPF WRR};
\end{axis}
\end{tikzpicture}
\caption{Aggregated UDP goodput with Turris Omnia.}
    \label{plot:aggreg-cpu}
\end{figure}

We use our lab (Figure \ref{subsec:bpf-perfs}) to study the impact on the forwarding performances
of the SRH encapsulation done in BPF and of the decapsulation performed by the kernel.
UDP flows are generated between the end hosts using \texttt{iperf3} with different
payload sizes at a 1~Gbps rate. 
The results are shown in Figure \ref{plot:aggreg-cpu}. 
The Turris Omnia is always the bottleneck. The decapsulation induces a 10\% overhead.
The eBPF WRR is running without the JIT compiler, because of a bug in
the current ARM32 implementation. 
As a consequence, the eBPF interpreter, which heavily consumes
CPU resources, is the bottleneck. The setup is however almost capable of
reaching the baseline performance for 1400-byte payloads, the 1.8 $\times$ speedup
factor provided by the JIT compiler, as demonstrated in Subsection \ref{subsec:bpf-perfs},
could be leveraged here with a functioning ARM32 implementation.

Our first experiments with TCP in this environment were a
disaster. Despite an aggregated bandwidth of 80~Mbps, the TCP
goodput reported by \texttt{nttcp} could only 
reach 3.8~Mbps. This low TCP performance is due to the difference
in delays over the two links that cause TCP reordering. 
Commercial solutions for hybrid access networks reorder the
out-of-sequence packets by using either sequence numbers 
in the GRE tunnels \cite{RFC8157} or Multipath TCP proxies 
\cite{bonaventure2016multipath} on the 
CPE and the aggregation box.


Instead of using sequence numbers as in
\cite{RFC8157}, we mitigate re-ordering
by delaying the link with the lowest latency. We extend our \texttt{End.DM} implementation
to handle two-way delay (TWD) measurements and deploy it on the
CPE. Instead of being decapsulated by \texttt{End.DM}, the TWD probes 
have as last segment the IPv6 address of the querier. 
A daemon running on the aggregation box sends TWD
measurements at regular intervals to the CPE on both links. The probes return,
and the daemon computes the difference of delays between the two
links. Our daemon then applies a \texttt{tc netem} queuing
discipline to delay the packets on the
fastest path using the difference between latencies that it computed.
This strategy does not fully prevent re-ordering, but still enables TCP flows
to attain acceptable aggregated goodputs on links with different latencies.



We then generated TCP connections using
\texttt{nttcp}. Thanks to the delay compensation, TCP could
efficiently utilize the two links. A single TCP connection reached
reach on average 68~Mbps while four parallel connections reached
70~Mbps. 



\subsection{Querying ECMP nexthops}

Our third example is en enhanced version of \texttt{traceroute}. 
Given the prevalence of Equal Cost Multipath (ECMP)
\cite{hopps2000analysis}, it becomes more and more difficult for
network operators to inspect routing problems with variants 
of \texttt{traceroute} \cite{augustin2006avoiding}.
Using \texttt{End.BPF}, we developed the \texttt{End.OAMP} SRv6 
eBPF function which, when triggered by a packet, performs a FIB 
lookup to query the ECMP nexthops for its destination address
and returns them to an address indicated in a TLV by the prober.
Our modified traceroute leverages if possible this function at 
each hop, and otherwise falls back to the legacy ICMP mechanism.
The eBPF function is written in 60 SLOC, whereas our custom helper
returning the ECMP nexthops for a given address required only 50 SLOC in the kernel.
This example underlines that, in order to extend the set of functionalities
accessible to eBPF programs, new helpers can easily be added to the kernel.


\section{Conclusion}

Network programmability is high on the wish list of many
network operators. In this paper, we propose, implement
and evaluate an extension of the Linux implementation of 
IPv6 Segment Routing that enables in-network programming. 
We provide an eBPF interface and a set of helper functions 
that enable them to write their own eBPF functions and attach 
them to specific SRv6 segments. Our measurements indicate
that our eBPF functions have a minimal overhead compared to their
static variants. Their main benefit is that they are generic. We
illustrate this with three very different use cases (delay
measurements, hybrid access networks and a multipath-aware
traceroute). 
Our eBPF extensions open new ways for network operators 
and researchers to implement in-network functions. 


\subsection*{Software artefacts}

We release our extensions to the Linux kernel as well as the eBPF
code developed for the different use cases under a GPL license:
\url{https://github.com/Zashas/Thesis-SRv6-BPF}.

\section*{Acknowledgements}

This work was partially supported by a Cisco URP grant and by CFWB within the ARC-SDN project.

\bibliographystyle{ACM-Reference-Format}
\bibliography{paper}

\end{document}